\documentclass[12pt]{article}

\usepackage{graphics}
\usepackage{epsfig}
\usepackage{cite}
\input epsf

\title{Prompt charmonia production and polarization at LHC in the NRQCD with $k_T$-factorization. \\Part II: $\chi_c$ mesons}
\author{S.P.~Baranov$^{1}$, A.V.~Lipatov$^{2,\,3}$, N.P.~Zotov$^2$}

\begin{document}

\maketitle

\begin{center}

{\it $^1$P.N.~Lebedev Physics Institute, 119991 Moscow, Russia}\\
{\it $^2$Skobeltsyn Institute of Nuclear Physics, Lomonosov Moscow State University, 119991 Moscow, Russia}\\
{\it $^3$Joint Institute for Nuclear Research, Dubna 141980, Moscow region, Russia}\\

\end{center}

\vspace{5mm}

\begin{center}

{\bf Abstract }

\end{center}

In the framework of $k_T$-factorization approach,
the production of prompt $\chi_c$ mesons in $pp$ collisions
at the LHC energies is studied. Our consideration is based on the 
off-shell amplitudes for hard partonic
subprocesses $g^*g^*\to\chi_{cJ}$
and non-relativistic QCD formalism for bound states.  
The transverse momentum dependent (unintegrated) gluon 
densities in a proton were derived from
Ciafaloni-Catani-Fiorani-Marchesini evolution equation or, alternatively, were
chosen in accordance with Kimber-Martin-Ryskin prescription.
Taking into account both color singlet and color octet contributions,
we deduce the corresponding non-perturbative long-distance matrix elements
from the fits to the latest ATLAS data on $\chi_{c1}$ and $\chi_{c2}$ 
transverse momentum distributions at $\sqrt s = 7$~TeV.
We find that these distributions at small and moderate $p_T$
are formed mainly by the color singlet components.
We successfully described the data on the differential cross sections
and relative production rates $\sigma(\chi_{c2})/\sigma(\chi_{c1})$
presented by the ATLAS, CMS and LHCb Collaborations. 
We find that the fit points to unequal wave functions 
of $\chi_{c1}$ and $\chi_{c2}$ states.

\vspace{1.0cm}

\noindent
PACS number(s): 12.38.-t, 13.20.Gd, 14.40.Pq

\newpage

\section{Introduction} \indent

Since it was first observed, charmonium production in hadronic collisions 
remains a subject of considerable theoretical and experimental interest. 
It provides a sensitive tool probing Quantum Chromodynamics (QCD) in both 
perturbative and non-perturbative regimes, as the production mechanism 
involves both short and long distance interactions. 
Two theoretical approaches for the non-perturbative part
are known in the literature: the 
color-singlet (CS) model\cite{1} and the color-octet (CO) model\cite{2}. 
As we have explained in our previous paper\cite{3},
none of the existing theoretical approaches is able to describe
all of the data in their integrity.
Our present study is a continuation of the work\cite{3}, 
where the prompt $\psi(2S)$ production and polarization at the LHC has been considered.
The motivation for the whole business has already been given 
there. 
Here we turn to the production of $P$-wave states. 
It is known that the feed-down contributions
from $\chi_c$ and $\psi(2S)$ states due to 
their radiative decays $\chi_c \to J/\psi + \gamma$ and
$\psi(2S) \to J/\psi + \gamma$
give a significant impact on the $J/\psi$ polarization\cite{4,5,6,7,8}.
These mechanisms constitute about $30$\% of the visible $J/\psi$ cross 
section at the LHC\cite{6,7,8}.
Therefore, a clear understanding of $\chi_c$ and $\psi(2S)$ production is a crucial 
component of any general description of $J/\psi$ production.
Another important issue concerns
the relative production rate 
$\sigma(\chi_{c2})/\sigma(\chi_{c1})$ at high transverse momenta.
This ratio is sensitive to the CS and CO mechanisms 
and can provide information complementary
to the study of the $S$-wave states\cite{9,10}.


Below we present a systematic analysis of the ATLAS\cite{11}, CMS\cite{12}
and LHCb\cite{13} data collected at $\sqrt s = 7$~TeV for $\chi_{c1}$ and
$\chi_{c2}$ the transverse momentum distributions and for the ratio of the
production rates $\sigma(\chi_{c2})/\sigma(\chi_{c1})$.

\section{Theoretical framework} \indent

Our consideration is based on the off-shell gluon-gluon fusion subprocess
that represents the true leading order in QCD:
\begin{equation}\label{gluglu}
  g^*(k_1)+g^*(k_2)\to c \bar{c} \to \chi_{cJ}(p),
\end{equation}

\noindent 
where the 4-momenta of all particles are indicated in 
parentheses.
In general, the charmed quark pair is produced in a state $^{2S+1}L_J^{(a)}$
with spin $S$, orbital angular momentum $L$, total angular momentum $J$ and 
color $a$, which can be either identical to the final charmonium quantum
numbers, as it is accepted in the CS model, or different 
from those. In the latter case, the $c\bar{c}$ pair transforms into physical
charmonium state by means of soft (non-perturbative) gluon radiation, as it
is considered in the formalism of non-relativistic QCD (NRQCD)\cite{14,15}.
The probability to form a given bound state is determined by the respective
non-perturbative long-distance matrix elements (NMEs), which are assumed to be
universal (process-independent), not depending on the charmonium momentum and
obeying certain hierarchy in powers of the relative charmed quarks velocity $v$.

The production of heavy $c\bar{c}$ pairs in hard partonic subprocess
is regarded as a purely perturbative stage and is considered 
in the framework of $k_T$-factorization approach\cite{16,17},
where studying the quarkonium production and polarization 
has a long history (see, for example,\cite{18,19,20,21,22,23,24,25,26,27,28} 
and references therein). 
A detailed description and discussion of the different aspects
of $k_T$-factorization can be found in reviews\cite{29}.
Here we see certain advantages in the fact that, even with the leading-order 
(LO) matrix elements for hard partonic subprocess, we can include a large 
piece of higher order QCD corrections (all NLO + NNLO + ... terms containing 
$\log 1/x$ enhancement) taking them into account in the form of transverse 
momentum dependent (TMD) gluon densities. The latter are obtained as numerical
solutions to Balitsky-Fadin-Kuraev-Lipatov (BFKL)\cite{30} 
or Catani-Ciafaloni-Fiorani-Marchesini (CCFM)\cite{31} evolution equations
and will be described below in more detail.

Summing over the initial gluon polarizations in (\ref{gluglu}) is done using the
spin density matrix 
$\overline{\epsilon^\mu\epsilon^\nu} = {\mathbf k}_T^{\mu}{\mathbf k}_T^{\nu}/|{\mathbf k}_T|^2$, 
where ${\mathbf k}_T$
is the component of the gluon momentum perpendicular to the proton beam direction\cite{16,17}.
In the collinear limit, when $|{\mathbf k}_T|\to 0$, this expression
converges to the ordinary $\overline{\epsilon^\mu\epsilon^\nu}=-g^{\mu\nu}/2$,
while for non-zero $|{\mathbf k}_T|$ gluon polarization vectors acquire an
admixture of longitudinal component. The evaluation of partonic amplitudes 
is straightforward and follow standard QCD Feynman rules in all other respects.
Our results for perturbative production amplitudes squared and summed
over polarization states agree with the ones\cite{32}. 

The formation of final state quarkonium from $c\bar{c}$ pair (with any quantum
numbers) is an essentially non-perturbative step.
At the LO in the relative quark velocity $v$, the $P$-wave mesons 
$\chi_{cJ}$ with $J = 0, 1$ or $2$ can be formed by a $c\bar c$ pair originally produced as color singlet 
$^3P_J^{(1)}$, or can evolve from an intermediate color octet $^3S_1^{(8)}$ state. 
The corresponding amplitudes can be obtained from an unspecified $c\bar c$ case by
applying the relevant projection operators\cite{1}:
\begin{equation}
  \Pi\left[^3S_1\right] =\hat\epsilon(S_z)\left( \hat p_c + m_c\right)/m^{1/2},
\end{equation}
\begin{equation}
  \Pi\left[^3P_J\right] =\left(\hat p_{\bar c} - m_c\right) \hat \epsilon(S_z) 
\left( \hat p_c + m_c\right)/m^{3/2},
\end{equation}

\noindent
where $m=2m_c$ is the mass of the considered $c\bar c$ state, $p_c$ and 
$p_{\bar c}$ are the four-momenta of the charmed quark and anti-quark,
$p_c=p/2+q$, $p_{\bar c}=p/2-q$, and $q$ is the the relative four-momentum 
of the quarks in the bound state.
States with various projections of the spin momentum onto the $z$ axis are 
represented by the polarization four-vector $\epsilon_\mu(S_z)$. 
The probability for the charmed quarks to form a meson depends on the real (for
color singlets) or fictituous (for color octets) bound state wave functions
$\Psi^{(a)}(q)$.
The corresponding NMEs are
related to the wave functions in the coordinate space ${\cal R}^{(a)}(x)$,
which are the Fourier transforms of $\Psi^{(a)}(q)$, and their derivatives\cite{2,14,15}:
\begin{equation}
 \langle{\cal O}\left[{^{2S + 1}L_J^{(a)}}\right]\rangle = 
 2 N_c (2J + 1) |{\cal R}^{(a)}(0)|^2 /4\pi,
\end{equation}

\noindent
for $S$-waves and
\begin{equation}
 \langle{\cal O}\left[{^{2S + 1}L_J^{(a)}}\right]\rangle = 
 6 N_c (2J + 1) |{\cal R}^{\prime\,(a)}|^2 /4\pi,
\end{equation}

\noindent
for $P$-waves. For more details the reader can address the
original papers\cite{1, 2,14,15} or our previous note\cite{3}. 
The CS NMEs
can be extracted from the measured $\chi_{c_2} \to \gamma \gamma$ decay width
or obtained from the potential models\cite{33,34,35,36}.
However, below we consider them as free parameters (as well as the CO NMEs) 
and determine from fits to the LHC data.

Finally, the $\chi_c$ meson production cross section is calculated as 
a convolution of the off-shell partonic cross sections and the TMD gluon 
densities in a proton:
\begin{equation}
 \displaystyle \sigma(pp\to\chi_{cJ}+X) =\int{2\pi\over x_1 x_2\,s\,F}\, 
 f_g(x_1,{\mathbf k}_{1T}^2,\mu^2) f_g(x_2,{\mathbf k}_{2T}^2,\mu^2)\,\atop
 \displaystyle \times \,|\bar {\cal A}(g^*+g^*\to\chi_{cJ})|^2\, 
 d{\mathbf k}_{1T}^2\,d{\mathbf k}_{2T}^2\,dy\,
 {d\phi_1 \over 2\pi} {d\phi_2 \over 2\pi},
\end{equation}

\noindent
where $f_g(x,{\mathbf k}_{T}^2,\mu^2)$ is the TMD gluon density, $y$ the 
rapidity of the produced $\chi_c$ meson and $\sqrt s$ the $pp$ center-of-mass energy.
The initial off-shell gluons carry longitudinal momentum fractions $x_1$ and 
$x_2$ (with respect to the parent protons) and non-zero transverse momenta 
${\mathbf k}_{1T}$ and ${\mathbf k}_{2T}$ 
oriented at the azimuthal angles $\phi_1$ and $\phi_2$. The off-shell flux 
factor $F$ is taken\footnote{Effect of the different forms of the flux factor on numerical
predictions has been studied in\cite{24}.} in accordance with the general definition\cite{37} 
as $F = 2\lambda^{1/2}(\hat s,k_1^2,k_2^2)$, where $\hat s = (k_1 + k_2)^2$.

In our numerical analysis we tried several sets of TMD gluon densities.
Two of them (A0\cite{38} and JH\cite{39}) have been obtained from CCFM 
equation where all input parameters have been fitted to the proton structure 
function $F_2(x, Q^2)$. Besides that, we used a parametrization obtained with
Kimber-Martin-Ryskin (KMR) prescription\cite{40} which provides a method
to construct TMD quark and gluon densities out of conventional (collinear) 
distributions. In that case, we used for the input the leading-order 
Martin-Stirling-Thorn-Watt (MSTW'2008) set\cite{41}.

The renormalization and factorization scales $\mu_R$ and $\mu_F$ were set 
to $\mu_R^2 = m^2 + {\mathbf p}_{T}^2$ and
$\mu_F^2 = \hat s + {\mathbf Q}_T^2$, where ${\mathbf Q}_T$ is the 
transverse momentum of the initial off-shell gluon pair. 
The choice of $\mu_R$ is rather standard for charmonium production, whereas 
the special choice of $\mu_F$ is connected with the CCFM evolution\cite{38,39}.
Following\cite{42}, we set the meson masses to 
$m_{\chi_{c1}}=3.51$~GeV and $m_{\chi_{c2}}=3.56$~GeV.
We use the LO formula for the running coupling constant $\alpha_s(\mu_R^2)$ 
with $n_f = 4$ quark flavours and $\Lambda_{\rm QCD} = 200$~MeV, so that 
$\alpha_s(M_Z^2) = 0.1232$.
The multidimensional integration has always been performed by the means 
of Monte Carlo technique using the routine \textsc{vegas}\cite{43}.
The full C$++$ code is available from the authors on 
request\footnote{lipatov@theory.sinp.msu.ru}.

\begin{table}
\begin{center}
\begin{tabular}{|c|c|c|c|}
\hline
  & & & \\
     & $|{\cal R}_{\chi_{c1}}^{\prime\,(1)}(0)|^2$/GeV$^5$ & 
       $|{\cal  R}_{\chi_{c2}}^{\prime\,(1)}(0)|^2$/GeV$^5$ & 
       $\langle {\cal O}^{\chi_{c0}}\left[^3S_1^{(8)}\right] \rangle$/GeV$^3$ \\
  & & & \\
\hline
  & & & \\
  A0 & $3.85 \times 10^{-1}$ & $6.18 \times 10^{-2}$ & $8.28 \times 10^{-5}$ \\
  & & & \\
  JH & $5.23 \times 10^{-1}$ & $9.05 \times 10^{-2}$ & $4.78 \times 10^{-5}$ \\
  & & & \\
  KMR & $3.07 \times 10^{-1}$ & $6.16 \times 10^{-2}$ & $1.40 \times 10^{-4}$ \\
  & & & \\
  \cite{9} & $7.50 \times 10^{-2}$ & $7.50 \times 10^{-2}$ & $2.01 \times 10^{-3}$ \\
  & & & \\
  \cite{10} & $3.50 \times 10^{-1}$ & $3.50 \times 10^{-1}$ & $4.40 \times 10^{-4}$ \\
  & & & \\
\hline
\end{tabular}
\caption{The NMEs for $\chi_c$ mesons and color singlet wave functions 
$|{\cal R}_{\chi_{c1}}^{\prime\,(1)}(0)|^2$ and $|{\cal R}_{\chi_{c2}}^{\prime\,(1)}(0)|^2$ 
extracted from the fit of the ATLAS data\cite{11}. 
The results obtained from the NLO NRQCD fits\cite{9,10} are shown for comparison.}
\label{table1}
\end{center}
\end{table}

The production of $\chi_c$ mesons is followed by their radiative decays.
Here we rely on the dominance of electric dipole transitions\footnote{The same
hypothesis has been used to study the production and polarization 
of $\Upsilon$ mesons at the Tevatron\cite{25}.}. The hypothesis 
of $E1$ dominance is supported by the data taken by the E835 Collaboration at 
the Tevatron\cite{44}. 
The corresponding decay amplitudes read\cite{45}
\begin{equation}
  {\cal A}(\chi_{c1}\to J/\psi +\gamma) \sim \epsilon^{\mu\nu\alpha\beta}k_\mu
  \epsilon_\nu^{(\chi_{c1})}\epsilon_\alpha^{(J/\psi)}\epsilon_\beta^{(\gamma)},
\end{equation}
\begin{equation}
  {\cal A}(\chi_{c2}\to J/\psi +\gamma) \sim
  p^\mu\epsilon^{\alpha \beta}_{(\chi_{c2})}\epsilon_\alpha^{(J/\psi)} 
 \left[ k_\mu \epsilon_\beta^{(\gamma)} - k_\beta\epsilon_\mu^{(\gamma)}\right],
\end{equation}
where $\epsilon^{\mu\nu\alpha\beta}$ is the fully antisymmetric Levi-Civita tensor,
$k$ is the final state photon 4-momentum, 
$\epsilon_\mu^{(\chi_{c1})}$, $\epsilon_\mu^{(J/\psi)}$ and $\epsilon_\mu^{(\gamma)}$
are the polarization vectors of the respective spin-one particles and 
$\epsilon_{\mu\nu}^{(\chi_{c2})}$ is the polarization tensor of spin-two $\chi_{c2}$ 
meson. The absolute decay rates were normalized to the known branchings 
$B(\chi_{c1}\to J/\psi +\gamma) = 0.344$ and
$B(\chi_{c2}\to J/\psi +\gamma) = 0.195$. 

\section{Numerical results} \indent

The whole set of NMEs was determined from fitting 
the transverse momentum distributions of $\chi_{c1}$ and $\chi_{c2}$ mesons measured
by the ATLAS Collaboration at $\sqrt s = 7$~TeV\cite{11}. The measurements 
were done at moderate and high transverse momenta $12 < p_T < 30$~GeV within the decay 
$J/\psi$ rapidity region $|y^{J/\psi}| < 0.75$, where the NRQCD formalism is believed 
to be reliable. The combined fit of $\chi_{c1}$ and $\chi_{c2}$ data was performed 
under the requirement that all NMEs be strictly positive. Following the 
suggestion\cite{27}, the $\chi_{c1}$ and $\chi_{c2}$ CS wave functions were 
treated as independent (not necessarily identical) parameters\footnote{The reasoning
refers to the fact that treating quarks as spinless particles in the potential 
models\cite{33,34,35,36} might be an oversimplification, and that radiative corrections
may be large.}.
The results of our fit are displayed in Table~1 for three different gluon 
distributions together with two sets of NMEs taken from the literature.
The calculated differential cross sections are presented in Figs.~1 and~2 as 
a functions of the $\chi_{cJ}$ and $J/\psi$ transverse momenta, respectively.
The ratio $\sigma(\chi_{c2})/\sigma(\chi_{c1})$ is shown in 
Fig.~3 in comparison with the recent LHC data\cite{11,12,13}.
Using the fitted values of NMEs from Table~1, we achieve good simulataneous
description of the measured $\chi_{c1}$ and $\chi_{c2}$ spectra 
and the ratio $\sigma(\chi_{c2})/\sigma(\chi_{c1})$ with each of the 
considered TMD gluon densities. 
We find that the $\chi_c$ production is dominated by the CS contributions,
that agrees with earlier conclusions\cite{20,21}.
However, a small CO admixture improves agreement with the LHC data at high 
transverse momenta (see Fig.~1).

The value of the CS wave function determined previously\cite{10}
from a combined fit to the Tevatron and LHC data is 
$|{\cal R}^{\prime\,(1)}(0)|^2 = 3.5 \times 10^{-1}$~GeV$^5$.
It differs
significantly from $|{\cal R}^{\prime\,(1)}(0)|^2 = 7.5 \times 10^{-2}$~GeV$^5$ 
obtained from the potential models\cite{33,34,35,36}.
The latter value is similar to the one extracted from the 
$\chi_{c2}\to\gamma\gamma$ decay width\cite{42}. Note that the authors 
of\cite{9,10}
assume equal values of the wave functions for $\chi_{c1}$ and $\chi_{c2}$
mesons. On the other hand, our fitting procedure leads to unequal values of the 
$\chi_{c1}$ and $\chi_{c2}$ CS wave functions. 
This qualitatively agrees with the suggestions\cite{27}
that the ratio of the wave functions has to be modified as
 $|{\cal R}_{\chi_{c1}}^{\prime\,(1)}(0)|^2/|{\cal R}_{\chi_{c2}}^{\prime\,(1)}(0)|^2 \sim 5$:$3$.
However, we find that the LHC data tend to support even larger ratio, namely 
 $|{\cal R}_{\chi_{c1}}^{\prime\,(1)}(0)|^2/|{\cal R}_{\chi_{c2}}^{\prime\,(1)}(0)|^2 \sim 5$:$1$.
Our fitted value of $|{\cal R}_{\chi_{c2}}^{\prime\,(1)}(0)|^2$
(but not $|{\cal R}_{\chi_{c1}}^{\prime\,(1)}(0)|^2$) is close
to the estimations based on
the potential models\cite{33,34,35,36} and two-photon decay width\cite{42}.


\section{Conclusions} \indent

We have considered prompt $\chi_c$ production in $pp$ collisions at the
energy $\sqrt s = 7$~TeV in the framework of the $k_T$-factorization approach
incorporated with non-relativistic QCD formalism.
Using the TMD gluon densities in a proton either derived from CCFM equation 
or constructed with Kimber-Martin-Ryskin method,
we extracted the corresponding nonperturbative color-singlet and color-octet
matrix elements from a combined fit to transverse momentum distributions 
of $\chi_{c1}$ and $\chi_{c2}$ mesons provided by the latest ATLAS 
measurements.
Using the fitted NMEs, we successfully described the data on the
relative production rates $\sigma(\chi_{c2})/\sigma(\chi_{c1})$
presented by the ATLAS, CMS and LHCb Collaborations. 
We find that the $\chi_c$ production is dominated by the CS contribbutions. 
However, an admixture of CO contributions improves the description of the data 
at high transverse momenta. Our interpretation of the LHC data supports the
idea of unequal values of the $\chi_{c1}$ and $\chi_{c2}$ CS wave functions.

\section{Acknowledgements} \indent 

The authors are grateful to H.~Jung
for very useful discussions and important remarks.
This research was supported by the FASI of Russian Federation
(grant NS-3042.2014.2).
We are also grateful to DESY Directorate for the
support in the framework of Moscow---DESY project on Monte-Carlo implementation for
HERA---LHC.

\newpage

\begin{figure}
\begin{center}
\epsfig{figure=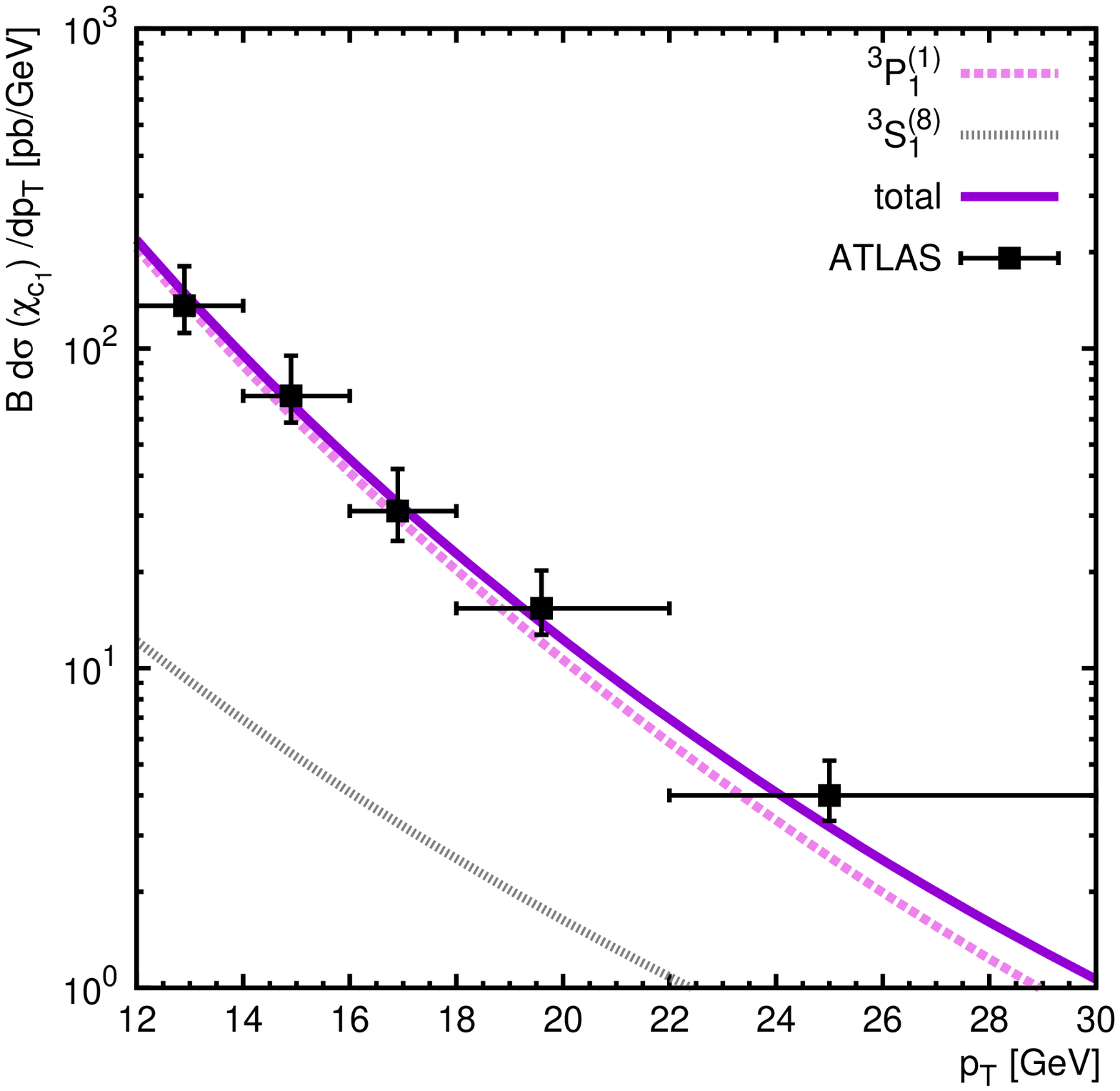, width = 6cm, angle = 270} 
\vspace{0.7cm} \hspace{-1cm}
\epsfig{figure=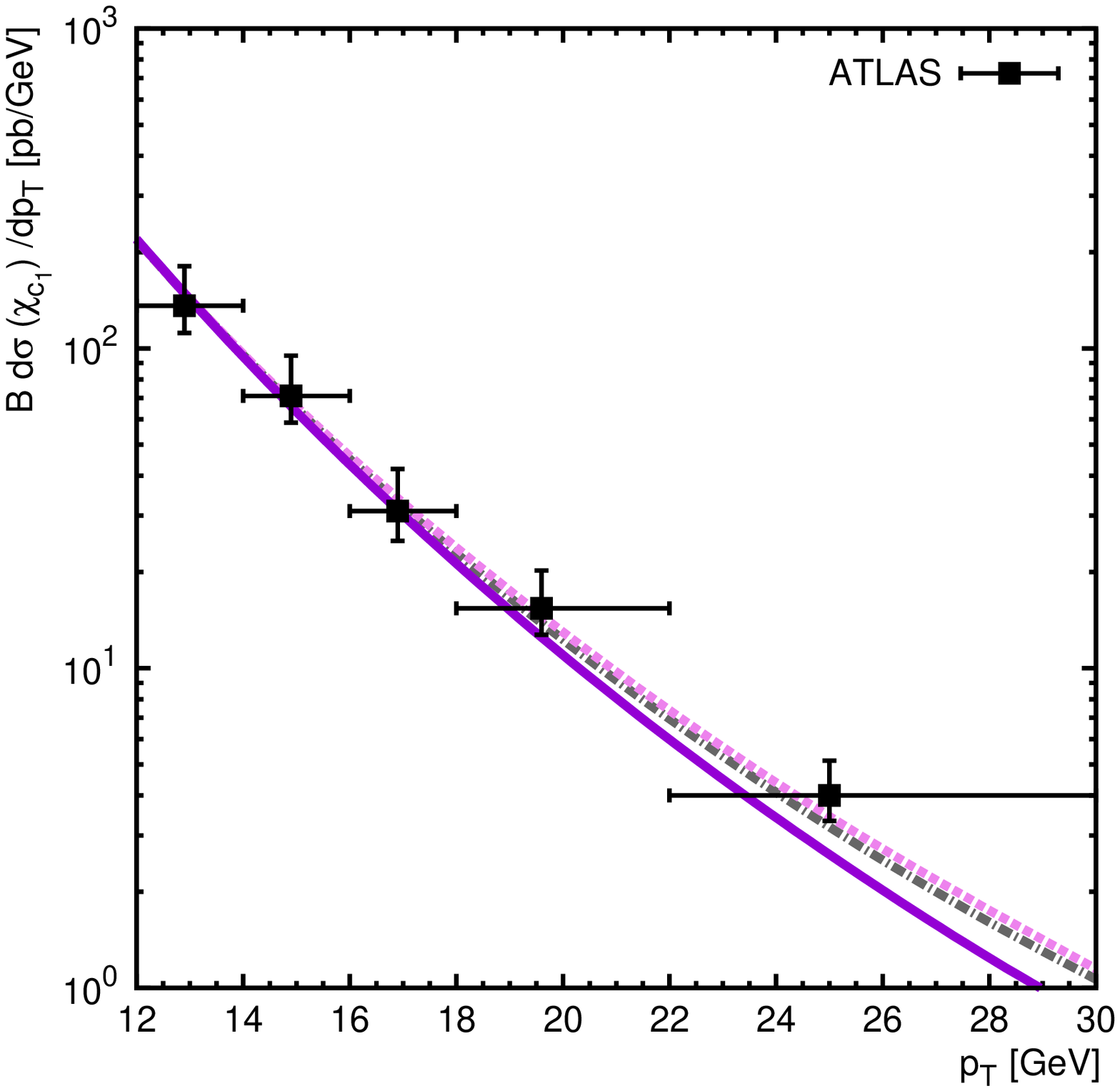, width = 6cm, angle = 270} 
\vspace{0.7cm}
\epsfig{figure=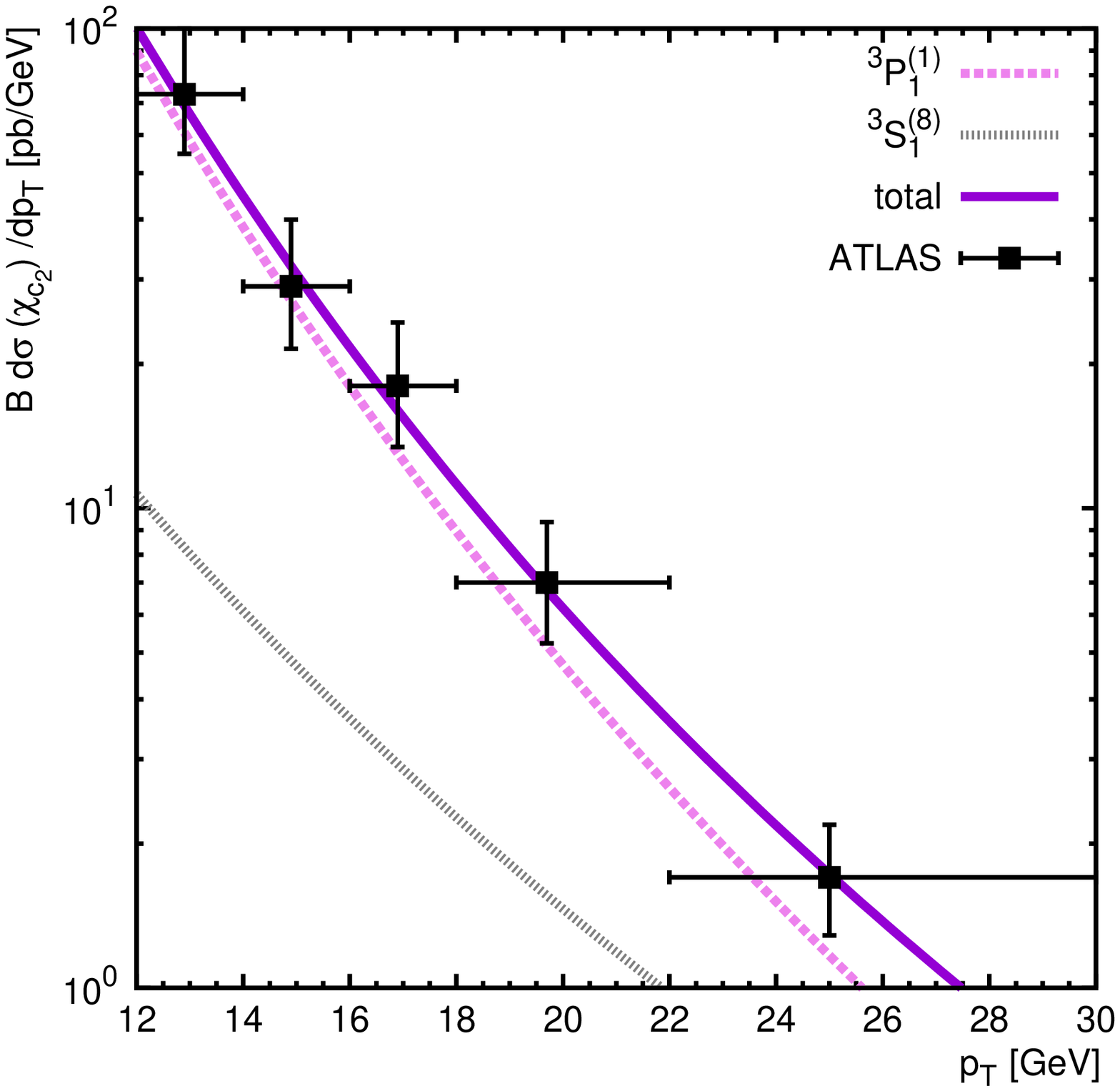, width = 6cm, angle = 270}
\hspace{-1cm}
\epsfig{figure=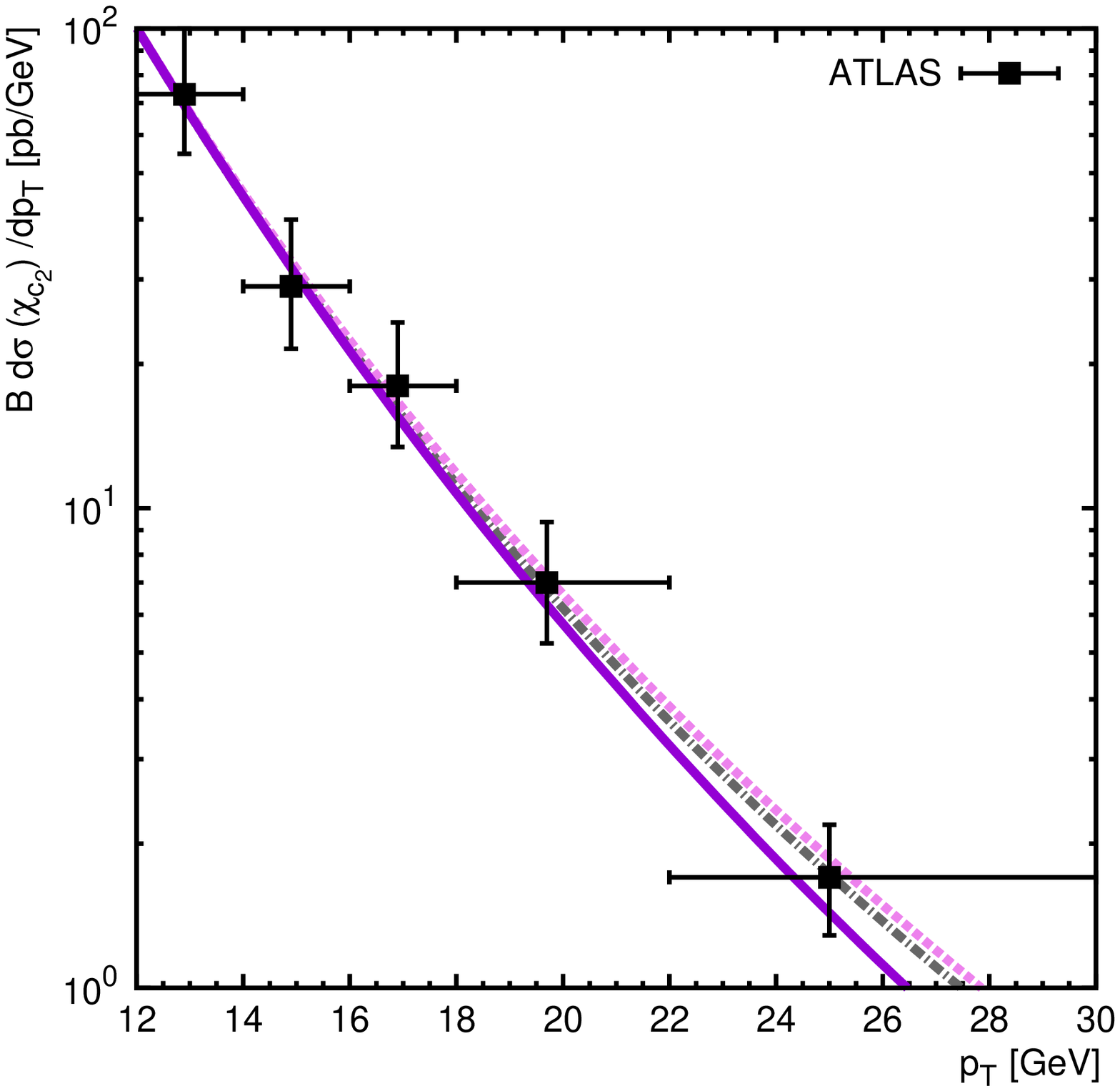, width = 6cm, angle = 270}
\caption{The prompt $\chi_c$ production at the LHC calculated
as a function of $\chi_c$ mesons transverse momenta at $\sqrt s = 7$~TeV.
Left panel: the dashed and dotted curves correspond to the 
color-singlet $^3P_J^{(1)}$ and color-octet 
$^3S_1^{(8)}$ contributions calculated with the KMR gluon density.
The solid curve represent the sum of CS and CO terms.
Right panel: the solid, dashed and dash-dotted curves correspond to the 
predictions obtained with the A0, JH and KMR gluon densities, 
respectively. The experimental data are from ATLAS\cite{11}.}
\label{fig1}
\end{center}
\end{figure}

\begin{figure}
\begin{center}
\epsfig{figure=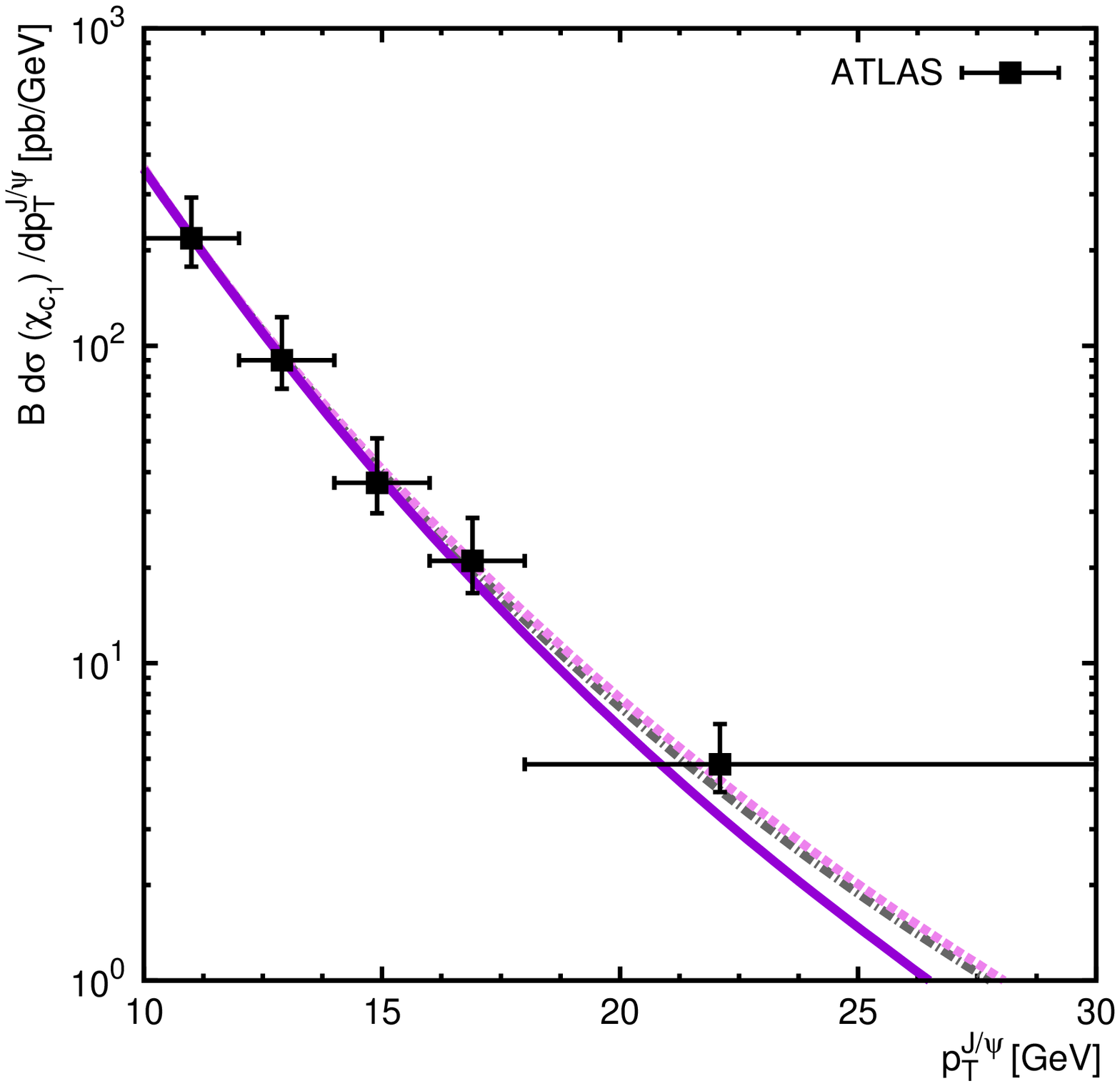, width = 6cm, angle = 270}
\hspace{-1cm}
\epsfig{figure=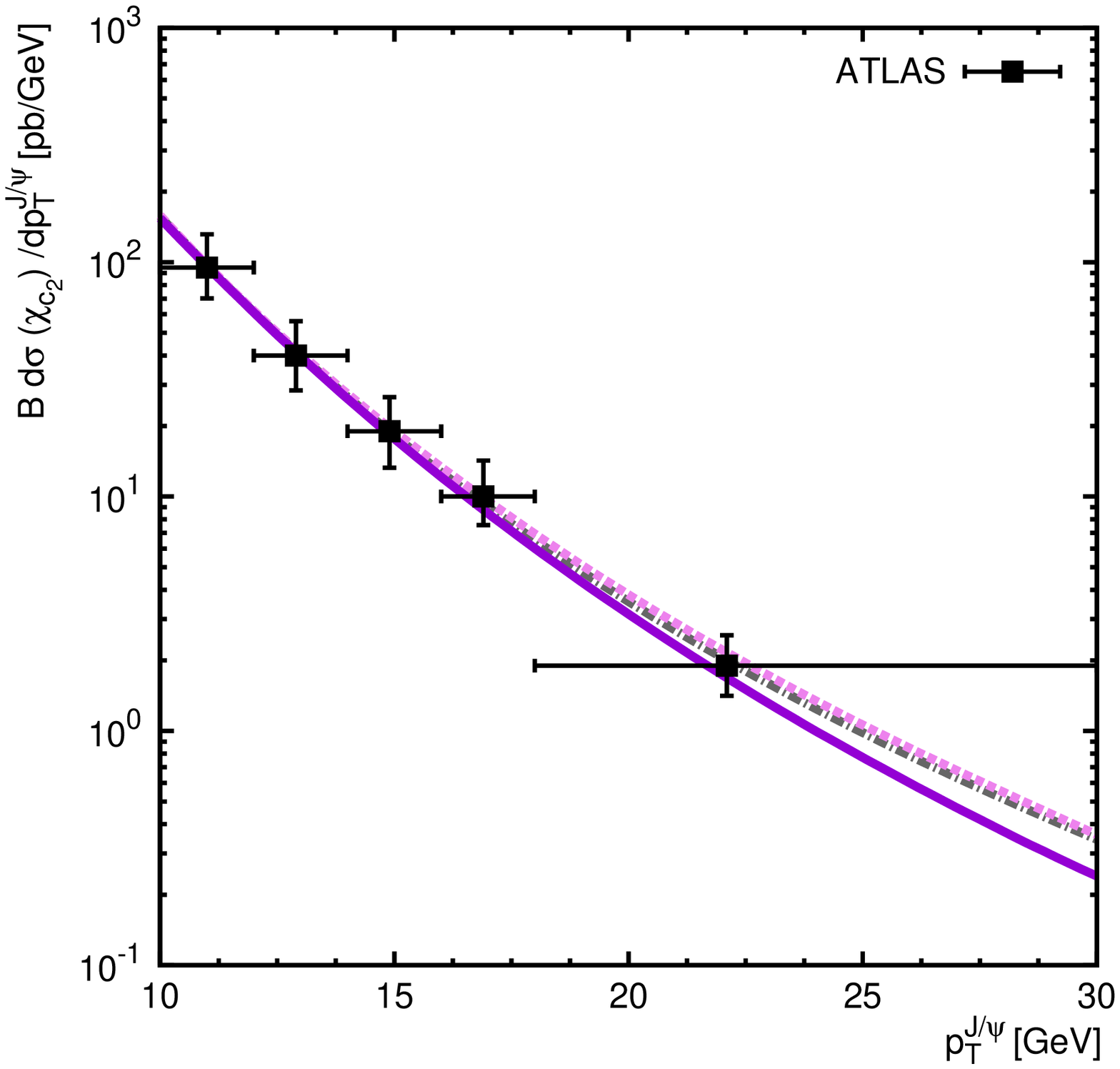, width = 6cm, angle = 270}
\caption{The prompt $\chi_c$ production at the LHC calculated
as a function of decay $J/\psi$ transverse momenta at $\sqrt s = 7$~TeV.
The solid, dashed and dash-dotted curves correspond to the 
predictions obtained with the A0, JH and KMR gluon densities, 
respectively. The experimental data are from ATLAS\cite{11}.}
\label{fig2}
\end{center}
\end{figure}

\begin{figure}
\begin{center}
\epsfig{figure=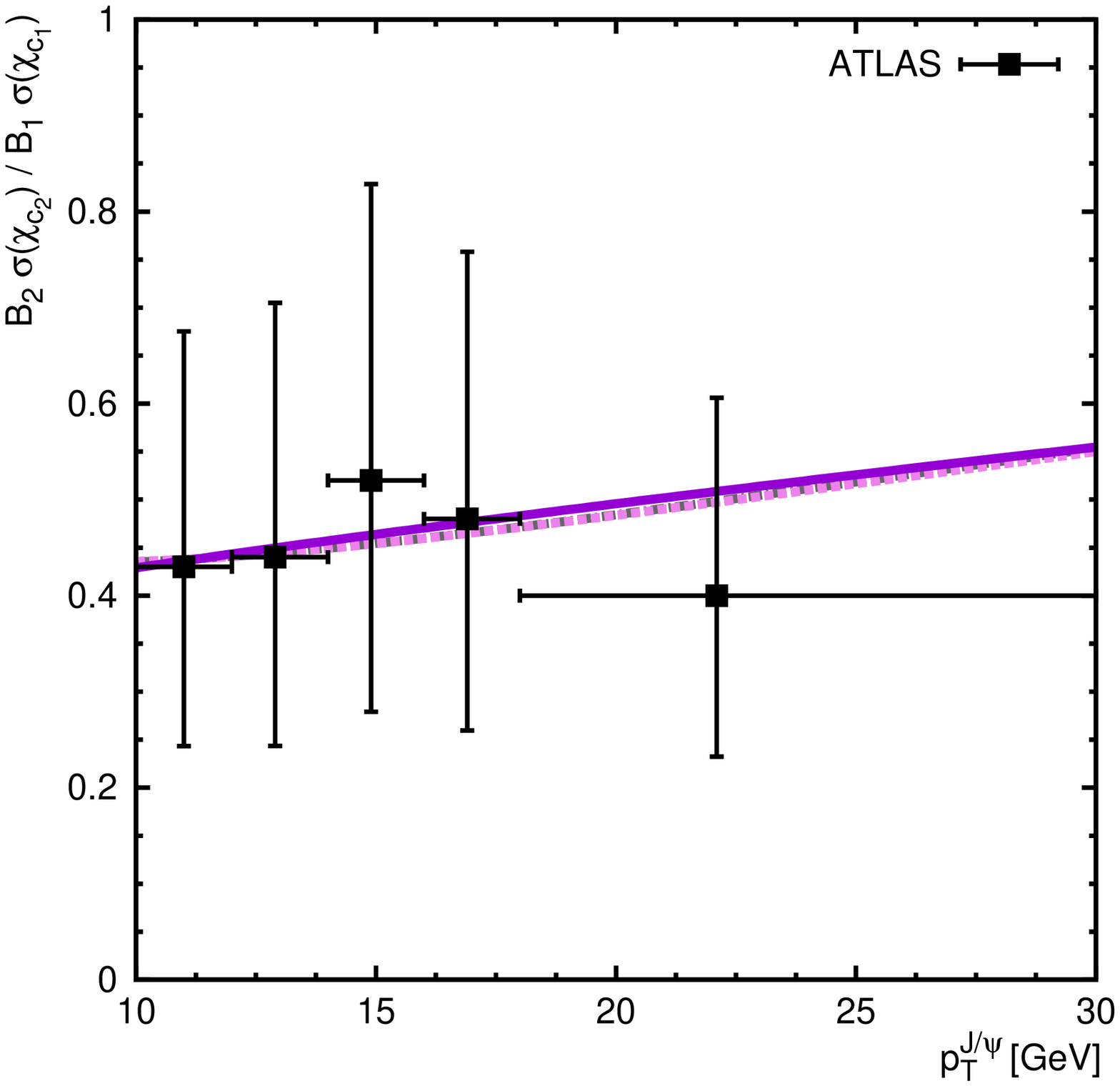, width = 6cm, angle = 270} 
\vspace{0.7cm} \hspace{-1cm}
\epsfig{figure=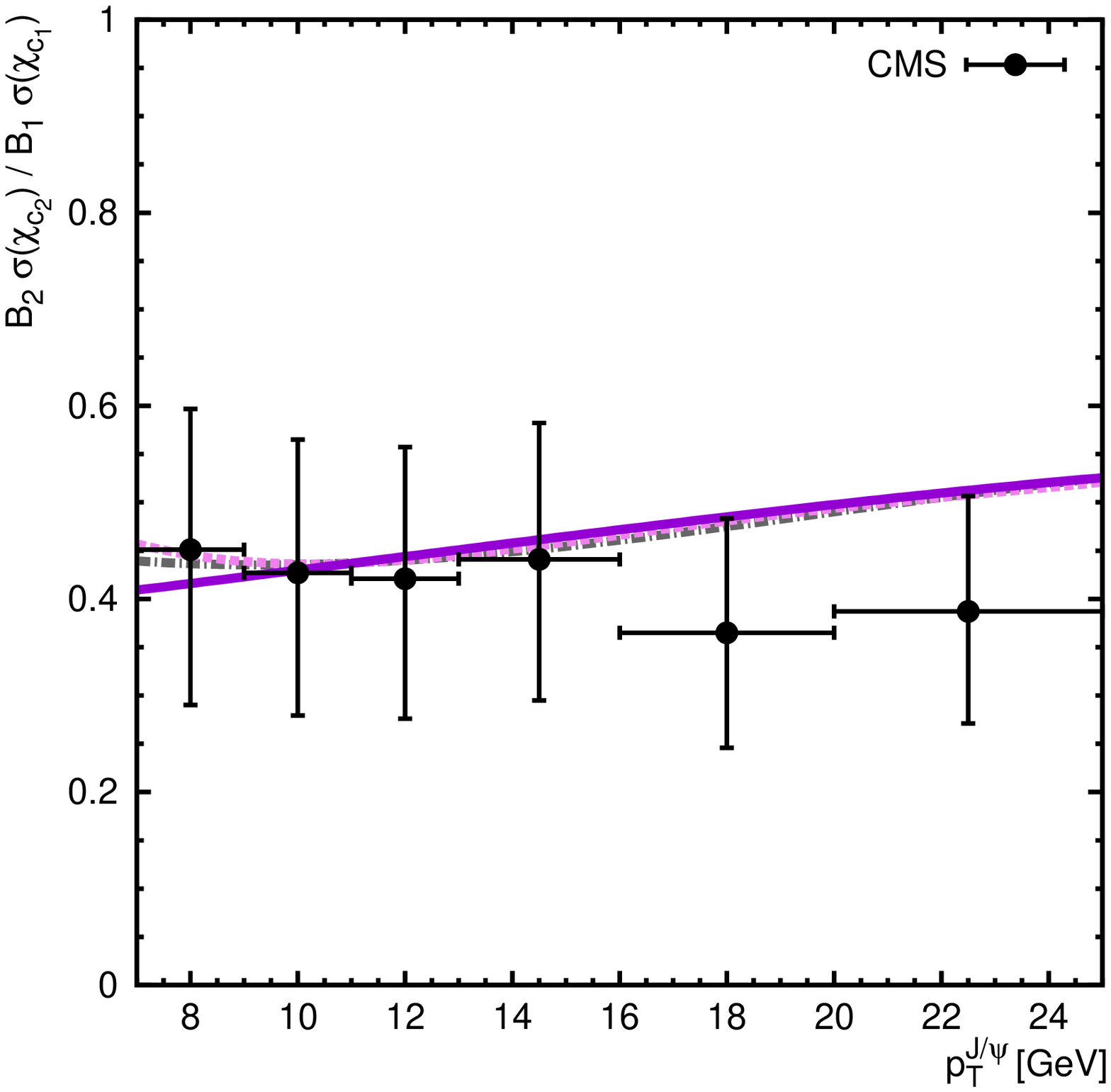, width = 6cm, angle = 270} 
\vspace{0.7cm}
\epsfig{figure=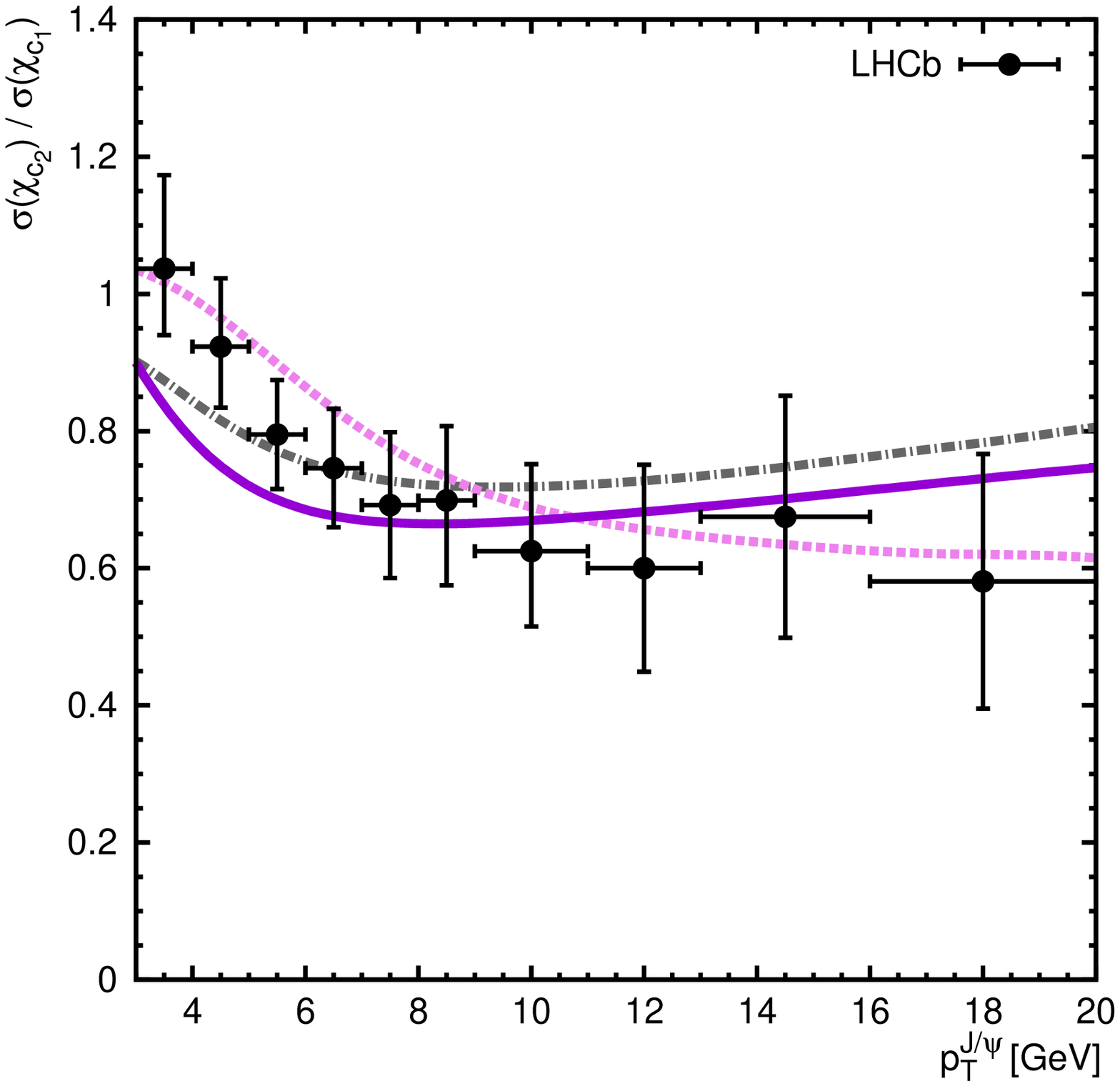, width = 6cm, angle = 270}
\caption{The relative production rate $\sigma(\chi_{c2})/\sigma(\chi_{c1})$ 
calculated as a function of $J/\psi$ meson transverse momenta 
at $\sqrt s = 7$~TeV. 
Notation of all curves is the same as in Fig.~2. 
The experimental data are from ATLAS\cite{11}, CMS\cite{12} and LHCb\cite{13}.}
\label{fig3}
\end{center}
\end{figure}


\begin{thebibliography}{55}

\bibitem{1} C.-H.~Chang, Nucl. Phys. B {\bf 172}, 425 (1980);\\
  E.L.~Berger, D.L.~Jones, Phys. Rev. D {\bf 23}, 1521 (1981);\\
  R.~Baier, R.~R\"uckl, Phys. Lett. B {\bf 102}, 364 (1981); \\
  S.S.~Gershtein, A.K.~Likhoded, S.R.~Slabospitsky, Sov. J. Nucl. Phys. {\bf 34}, 128 (1981).
\bibitem{2} E.~Braaten, S.~Fleming, Phys. Rev. Lett. {\bf 74}, 3327 (1995).
\bibitem{3} S.P.~Baranov, A.V.~Lipatov, N.P.~Zotov, Eur. Phys. J. C {\bf 75}, 455 (2015).
\bibitem{4} CDF Collaboration, Phys. Rev. Lett. {\bf 99}, 132001 (2007).
\bibitem{5} CDF Collaboration, Phys. Rev. Lett. {\bf 108}, 151802 (2012).
\bibitem{6} CMS Collaboration, Phys. Lett. B {\bf 727}, 381 (2013). 
\bibitem{7} LHCb Collaboration, Eur. Phys. J. C {\bf 74}, 2872 (2014). 
\bibitem{8} CMS Collaboration, Phys. Rev. Lett. {\bf 110}, 081802 (2013).
\bibitem{9} H.-F.~Zhang, L.~Yu, S.-X.~Zhang, L.~Jia, arXiv:1410.4032 [hep-ph].
\bibitem{10} A.K.~Likhoded, A.V.~Luchinsky, S.V.~Poslavsky, Phys. Rev. D {\bf 90}, 074021 (2014). 
\bibitem{11} ATLAS Collaboration, JHEP {\bf 07}, 154 (2014). 
\bibitem{12} CMS Collaboration, Eur. Phys. J. C {\bf 72}, 2251 (2012). 
\bibitem{13} LHCb Collaboration, JHEP {\bf 10}, 115 (2013). 
\bibitem{14} G.~Bodwin, E.~Braaten, G.~Lepage, Phys. Rev. D {\bf 51}, 1125 (1995).
\bibitem{15} P.~Cho, A.K.~Leibovich, Phys. Rev. D {\bf 53}, 150 (1996); Phys. Rev. D {\bf 53}, 6203 (1996).
\bibitem{16} L.V.~Gribov, E.M.~Levin, M.G.~Ryskin, Phys. Rep. {\bf 100}, 1 (1983);\\
  E.M.~Levin, M.G.~Ryskin, Yu.M.~Shabelsky, A.G.~Shuvaev, Sov. J. Nucl. Phys. {\bf 53}, 657 (1991).
\bibitem{17} S.~Catani, M.~Ciafaloni, F.~Hautmann, Nucl. Phys. B {\bf 366}, 135 (1991);\\
  J.C.~Collins, R.K.~Ellis, Nucl. Phys. B {\bf 360}, 3 (1991).
\bibitem{18} S.P.~Baranov, Phys. Lett. B {\bf 428}, 377 (1998).
\bibitem{19} F.~Yan, K.T.~Chao, Phys. Rev. Lett. {\bf 87}, 022002 (2001).
\bibitem{20} Ph.~H\"agler, R.~Kirschner, A.~Sh\"afer, L.~Szymanowski, O.V.~Teryaev, Phys. Rev. D {\bf 63}, 077501 (2001).
\bibitem{21} S.P.~Baranov, Phys. Rev. D {\bf 66}, 114003 (2002).
\bibitem{22} A.V.~Lipatov, N.P.~Zotov, Eur. Phys. J. C {\bf 27}, 87 (2003).
\bibitem{23} S.P.~Baranov, N.P.~Zotov, J. Phys. G {\bf 29}, 1395 (2003).
\bibitem{24} S.P.~Baranov, A.~Szczurek, Phys. Rev. D {\bf 77}, 054016 (2008).
\bibitem{25} S.P.~Baranov, N.P.~Zotov, JETP Lett. {\bf 88}, 711 (2008).
\bibitem{26} S.P.~Baranov, A.V.~Lipatov, N.P.~Zotov, Eur. Phys. J. C {\bf 71}, 1631 (2011).
\bibitem{27} S.P.~Baranov, Phys. Rev. D {\bf 83}, 034035 (2011).
\bibitem{28} S.P.~Baranov, A.V.~Lipatov, N.P.~Zotov, Phys. Rev. D {\bf 85}, 014034 (2012).
\bibitem{29} B.~Andersson {\sl et al.} (Small-$x$ Collaboration), Eur. Phys. J. C {\bf 25}, 77 (2002);\\
  J.~Andersen {\sl et al.} (Small-$x$ Collaboration), Eur. Phys. J. C {\bf 35}, 67 (2004);\\
  J.~Andersen {\sl et al.} (Small-$x$ Collaboration), Eur. Phys. J. C {\bf 48}, 53 (2006).
\bibitem{30} E.A.~Kuraev, L.N.~Lipatov, V.S.~Fadin, Sov. Phys. JETP {\bf 44}, 443 (1976);\\
  E.A.~Kuraev, L.N.~Lipatov, V.S.~Fadin, Sov. Phys. JETP {\bf 45}, 199 (1977);\\
  I.I.~Balitsky, L.N.~Lipatov, Sov. J. Nucl. Phys. {\bf 28}, 822 (1978).  
\bibitem{31} M.~Ciafaloni, Nucl. Phys. B {\bf 296}, 49 (1988);\\
  S.~Catani, F.~Fiorani, G.~Marchesini, Phys. Lett. B {\bf 234}, 339 (1990);\\
  S.~Catani, F.~Fiorani, G.~Marchesini, Nucl. Phys. B {\bf 336}, 18 (1990);\\
  G.~Marchesini, Nucl. Phys. B {\bf 445}, 49 (1995). 
\bibitem{32} B.A.~Kniehl, D.V.~Vasin, V.A.~Saleev, Phys. Rev. D {\bf 73}, 074022 (2006);\\
  V.A.~Saleev, M.A.~Nefedov, A.V.~Shipilova, Phys. Rev. D {\bf 85}, 074013 (2012).
\bibitem{33} C.R.~Munz, Nucl. Phys. A {\bf 609}, 364 (1996).
\bibitem{34} D.~Ebert, R.N.~Faustov, V.O.~Galkin, Mod. Phys. Lett. A {\bf 18}, 601 (2003).
\bibitem{35} G.-L.~Wang, Phys. Lett. B {\bf 674}, 172 (2009).
\bibitem{36} B.-Q.~Li, K.-T.~Chao, Commun. Theor. Phys. {\bf 52}, 653 (2009).
\bibitem{37} E.~Bycling, K.~Kajantie, Particle Kinematics, John Wiley and Sons (1973).
\bibitem{38} H.~Jung, arXiv:hep-ph/0411287.
\bibitem{39} F.~Hautmann, H.~Jung, Nucl. Phys. B {\bf 883}, 1 (2014).
\bibitem{40} M.A.~Kimber, A.D.~Martin, M.G.~Ryskin, Phys. Rev. D {\bf 63}, 114027 (2001);\\
  G.~Watt, A.D.~Martin, M.G.~Ryskin, Eur. Phys. J. C {\bf 31}, 73 (2003).
\bibitem{41} A.D.~Martin, W.J.~Stirling, R.S.~Thorne, G.~Watt, Eur. Phys. J. C {\bf 63}, 189 (2009).
\bibitem{42} PDG Collaboration, Chin. Phys. C {\bf 38}, 090001 (2014).
\bibitem{43} G.P.~Lepage, J. Comput. Phys. {\bf 27}, 192 (1978).
\bibitem{44} E835 Collaboration, Phys. Rev. D {\bf 65}, 052002 (2002).
\bibitem{45} P.~Cho, M.~Wise, S.~Trivedi, Phys. Rev. D {\bf 51}, R2039 (1995).

\end{thebibliography}
\end{document}